\documentstyle[aps,twocolumn,epsf]{revtex}

\newcommand{\gl}[1]{(\ref{#1})}

\def\simge{\mathrel{%
   \rlap{\raise 0.511ex \hbox{$>$}}{\lower 0.511ex \hbox{$\sim$}}}}
\def\simle{\mathrel{
   \rlap{\raise 0.511ex \hbox{$<$}}{\lower 0.511ex \hbox{$\sim$}}}}

\begin{document}

\title{ Gluon plasmon frequency near the light-cone}
 
\author{Fritjof Flechsig\footnote{Supported by 
                   Deutsche Forschungsgemeinschaft (DFG)
                   under grant no.~Schu 1045/1-2.}
       } 

\address{ Institut f\"ur Theoretische Physik, Universit\"at Hannover,\\ 
          Appelstr.~2, D-30167 Hannover, Germany\\
          e-mail: flechsig@itp.uni-hannover.de 
        } 

\date{September 1998}

\maketitle

\begin{abstract}

Thermal perturbation theory based on the resummation scheme by Braaten
and Pisarski suffers from unscreened collinear singularities whenever outer
momenta become light-like. A recently proposed improvement of the hard
thermal loops by an additional resummation of an asymptotic mass promises
to solve this problem. Here we present a detailed investigation of the
next-to-leading order plasmon self-energy. It demonstrates the applicability
and consistency of the improved scheme.

\end{abstract}

%
%

\section{Introduction}

The resummation scheme by Braaten and Pisarski \cite{BP} is well established 
in high temperature field theory. Since its first announcement in 1989 it has 
been applied to a vast variety of physical problems \cite{reviews}, where the 
calculation of the  gluon damping rate was the most spectacular one
\cite{BP-dampf}. To obtain consistent results in a perturbative calculation one
has to distinguish two momentum scales. On the hard scale, where momenta are 
$\sim T$ ($T=$ temperature), the conventional perturbation series is valid. 
On the soft scale, where the momenta are $\sim gT$ ($g$ is the small
coupling), one has to resum the so called 'hard thermal loops' (HTL). The 
resulting effective vertices and propagators are screening formerly 
untreatable infrared singularities. This leads to finite and gauge invariant 
outcomings for physical quantities.

In some cases, however, it turned out that the HTL were insufficient to
screen all infinities. For example, the damping rate of dynamical gluons and
the screening length in a gluon plasma suffer from the lack of a thermal 
magnetic mass \cite{FRoSch}, which results in an infrared singularity. Another
shortcoming  was introduced by the HTL itself. Whenever outer momenta are soft
and lightlike, the HTL cause collinear singularities \cite{srppr,sed,FleSch}. 
Nevertheless, in a medium like the gluon plasma there are no long range forces
since medium effects care for the screening. Thus, such singularities should
be spurious.

The Braaten-Pisarski scheme can be understood in the context of a
renormalisation group treatment \cite{B-renorm}. The effective propagators and
vertices are included in an effective action \cite{eff} which is the
result of integrating out the hard momentum scale. It was shown in \cite{sed},
however, that in this action for lightlike outer momenta the hard modes have 
not been integrated out completely. One has to take into account a certain
asymptotic thermal mass. An improved HTL resummation scheme has been proposed
recently in \cite{eFFacT} which includes this mass and is free from
collinear singularities while retaining the structure and simplicity of 
the action \cite{eff}. In this letter we will test the new resummation scheme
for consistency. Its applicability will be shown by a simple check of the 
next-to-leading order plasmon frequency.

The paper starts with a collection of notations (sect.~2).
The next two sections review a few facts on the plasmon-frequency at 
next-to-leading order (sect.~3) and on the resummation of asymptotic
masses (sect.~4). In section 5 we derive two consistency conditions
which the plasmon frequency must obey and in section 6 we test these conditions
by explicitly calculating the plasmon frequency near the light-cone.
It will be seen that the conditions are fulfilled. Conclusions are given in 
section 7.

\section{Notations}

Consider a system of gluons in thermal equilibrium at high temperature $T$. It
is described by the Lagrangian
$
 {\cal L}=- {1\over4} F_{\mu\nu}^aF^{{\mu\nu} a}
          - \frac{1}{2\alpha}\left(\partial^\mu A_\mu^a\right)
          + {\cal L}_{\rm ghost}
$
where $F_{\mu\nu}^a$ is the non-abelian $SU(N)$ field tensor. The coupling $g$ 
is small. For simplicity we do not allow for quarks. In the imaginary time
formalism each four momentum reads $Q=(Q_o,\vec q)$. $Q_o=i\omega_n=2\pi i nT$
are the Matsubara frequencies. We use the metric $g_{\mu\nu}=(+---)$ and a
covariant gauge-fixing with parameter $\alpha$.

The spectra of physical exitations in the plasma are defined by the poles of 
the response function $\chi(\omega,q) = G(Q_o\to\omega+i\eta,q)$, ($\eta\to 
+0$), where $G$ is the gluon propagator. The HTL resummed leading order of $G$
at soft momentum reads
\begin{eqnarray}
  {}^*G_{\mu\nu}(P)
   &=& A_{\mu\nu}(P)\Delta_t(P) + B_{\mu\nu}(P)\Delta_l(P) \nonumber\\
   && + \alpha D_{\mu\nu}(P)\Delta_o(P)
\label{eff100}
\end{eqnarray}
with
$\Delta_{t,\ell}:=[P^2-\Pi_{t,\ell}]^{-1}$,
$\Delta_o=1/P^2$,
$\Pi_t:=\frac{1}{2}{\rm Tr}\,(A\Pi)$ and
$\Pi_\ell:={\rm Tr}\,(B\Pi)$.
The Lorentz-matrices are taken from a basis, see eg. \cite{nt}, where
$A$ is the transversal and $B$ is the longitudinal projector.

The propagator has two physical poles \cite{LandsReb}. We will concentrate on
the longitudinal mode: the solution $\omega_\ell(q)$ of
\begin{eqnarray}
\omega_\ell^2-q^2&=&\Pi_\ell(\omega_\ell,q) \;, \nonumber\\
\Pi_\ell(Q)&=&-3m^2{Q^2\over q^2}
 \left\{ 1 - {Q_o\over 2q}\ln\left({Q_o+q\over Q_o-q}\right) \right\}\;,
\label{eff120}
\end{eqnarray}
is the frequency of the well known plasmon \cite{KlimWel}. $\Pi_\ell$ is the 
longitudinal part of the polarization function $\Pi_{\mu\nu}$ in HTL 
approximation. It is the sum of one-loop diagrams where the inner momentum 
$P$ is hard, 
\begin{equation}\label{eff130}
  \Pi_{\mu\nu}(Q)=2g^2N{\textstyle\sum}\!\!\!\!\!\!\int\nolimits_P
  \Delta_o^-\Delta_o\Big\{-P^2g_{\mu\nu}+2P_\mu P_\nu \Big\}
\;.
\end{equation}
The summation symbol is defined by 
$ {\scriptstyle\Sigma}\!\!\!\!\int\nolimits_P 
  := \int\!\frac{d^3p}{(2\pi)^3} T\sum_n$
where $n$ runs over all Matsubara frequencies. 
Throughout the paper $Q$ is the external momentum, $P$ is summed over and
$K$ is the difference $K=Q-P$. A superscript 'minus' refers to the 
transformation $P\to K$, e.g. $\Delta_o^- = 1/K^2$.

As \gl{eff120} shows, the sumintegral \gl{eff130} can be performed 
analytically. But to obtain the Braaten-Pisarski effective action one 
conveniently starts from
\begin{eqnarray}\label{eff140}
 \Pi_{\mu\nu}(Q)
 &=& 3m^2\left( U_\mu U_\nu-\int_\Omega\frac{(UQ)}{(YQ)}Y_\mu Y_\nu \right)\\
 &&\mbox{with}\quad 
Y=(1,\vec e)\quad,\quad
m^2={g^2NT^2\over 9} \;. \nonumber
\end{eqnarray}
The angular integral $\int_\Omega:={1\over 4\pi}\int d^2\Omega$ 
averages over the directions of the unit-vector  $\vec e$. 
Using this notation, the effective vertices obtain
a pleasant form. They split up into two parts, e.g.
\begin{equation}\label{eff150}
  {}^*\Gamma^{(3)}_{{\mu\nu}\sigma}(Q,P,K) 
   =  \Gamma^{\rm tree}_{{\mu\nu}\sigma} (Q,P,K)
    + \delta\Gamma^{(3)}_{{\mu\nu}\sigma}(Q,P,K) \;,
\end{equation}
where $\Gamma^{\rm tree}_{{\mu\nu}\sigma}$ is the tree--level and 
$\delta\Gamma^{(3)}_{{\mu\nu}\sigma}$ the HTL part of the 3-gluon-vertex,
\begin{eqnarray}
 \delta\Gamma^{(3)}_{{\mu\nu}\rho} 
  & = & 3m^2\int_\Omega {Y_\mu Y_\nu Y_\rho\over YQ}
        \left\{ \frac{P_o}{YP}-\frac{K_o}{YK} \right\} \;,
\label{eff160}\\
 \delta\Gamma^{(4)}_{{\mu\nu}\rho\lambda} 
  & = & 6m^2\int_\Omega {Y_\mu Y_\nu Y_\rho Y_\lambda\over (YQ)^2}
       \left\{ \frac{P_o}{YP}-\frac{K_o}{YK} \right\}\;. 
\label{eff170}
\end{eqnarray}
For the derivation of the HTL effective action from the above we refer to
\cite{eff,eFFacT}.

\section{Next to leading order of the Plasmon Frequency}

The next to leading order of the plasmon frequency has been extensively studied
in \cite{BP-dampf,FleSch,nt}. Up to next to leading order the plasmon is the 
solution of the equation
\begin{equation}\label{eff200}
 \omega^2-q^2=\overline\Pi_\ell(\Omega+i\eta,q)
\quad{\rm or}\quad
-q^2=\overline\Pi_{oo}(\Omega+i\eta,q) 
\end{equation}
since $\Pi_{\mu\nu}$ is transversal up to this order \cite{Kunst,FleSch}. The 
complex frequency $\Omega=\omega_\ell\, +\, \delta\omega_\ell -i\gamma_\ell$ 
is the sum of the plasmon frequency at leading order, its next-to-leading
correction and the plasmon damping,  while 
$\overline\Pi_{oo} = \Pi_{oo}+\delta\Pi_{oo}$ 
is the polarization function including next-to-leading order. We expand 
\gl{eff200} and find for the physical quantities
\begin{eqnarray}
\delta\omega_\ell(q) 
 &=&  - {\Re e\,\delta\Pi_{oo}(\omega_\ell+i\eta,q) \over 
       \left.\partial_\omega \Pi_{oo}(\omega,q)\right |_{\omega=\omega_\ell} }
\;,\label{eff210} \\
\gamma_\ell(q)
 &=&  {\Im m\,\delta\Pi_{oo}(\omega_\ell+i\eta,q) \over   
       \left.\partial_\omega \Pi_{oo}(\omega,q)\right |_{\omega=\omega_\ell} }
\;.\label{eff215}
\end{eqnarray}
Thus, to calculate $\delta\omega_\ell(q)$ or $\gamma_\ell(q)$ one has to 
evaluate $\delta\Pi_{oo}$ and $\Pi_{oo}$ on the leading order mass-shell
$\omega=\omega_\ell(q)$.
\begin{figure}
\begin{center}
 \parbox{5cm}{
  \epsfxsize 5cm
  \epsfbox{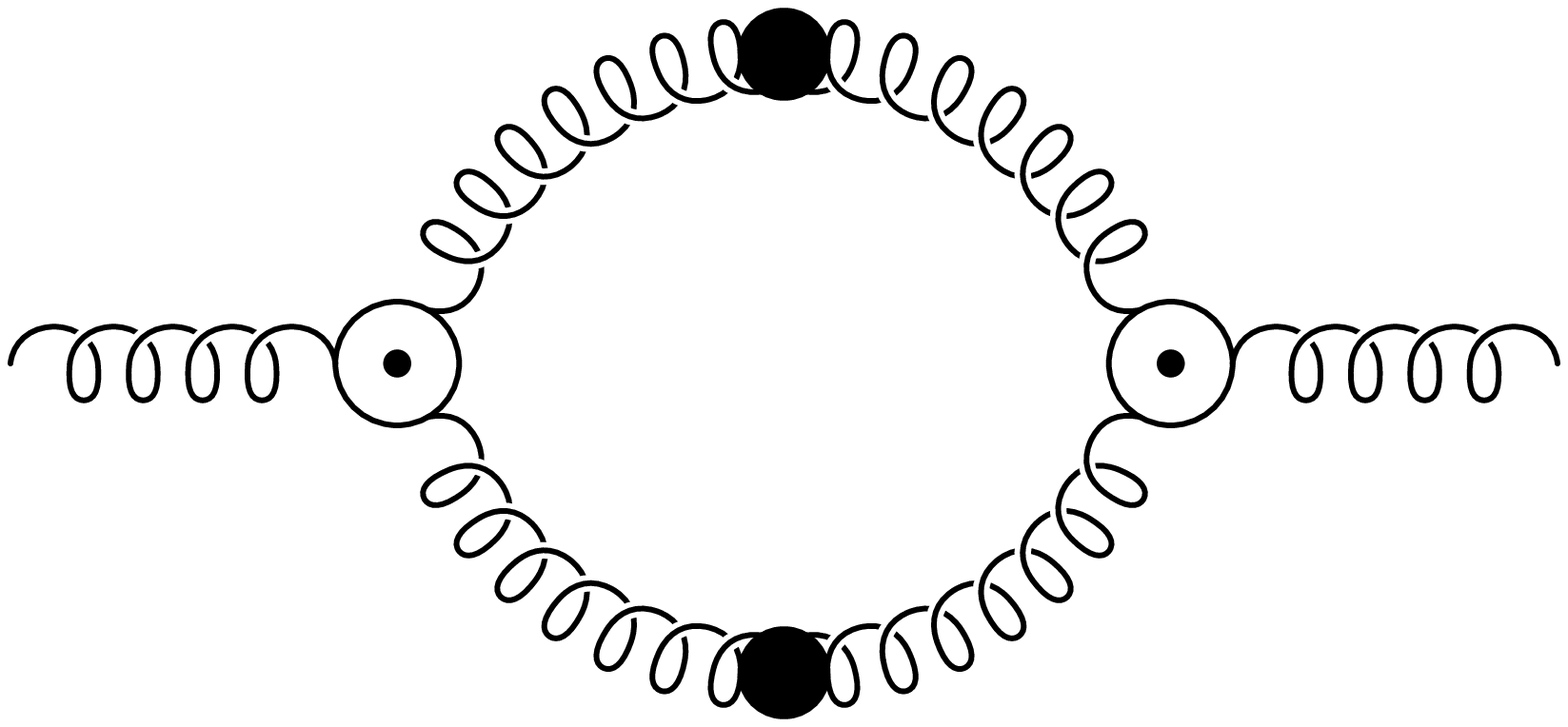}
 }
 \parbox{3cm}{
  \epsfxsize 3cm
  \epsfbox{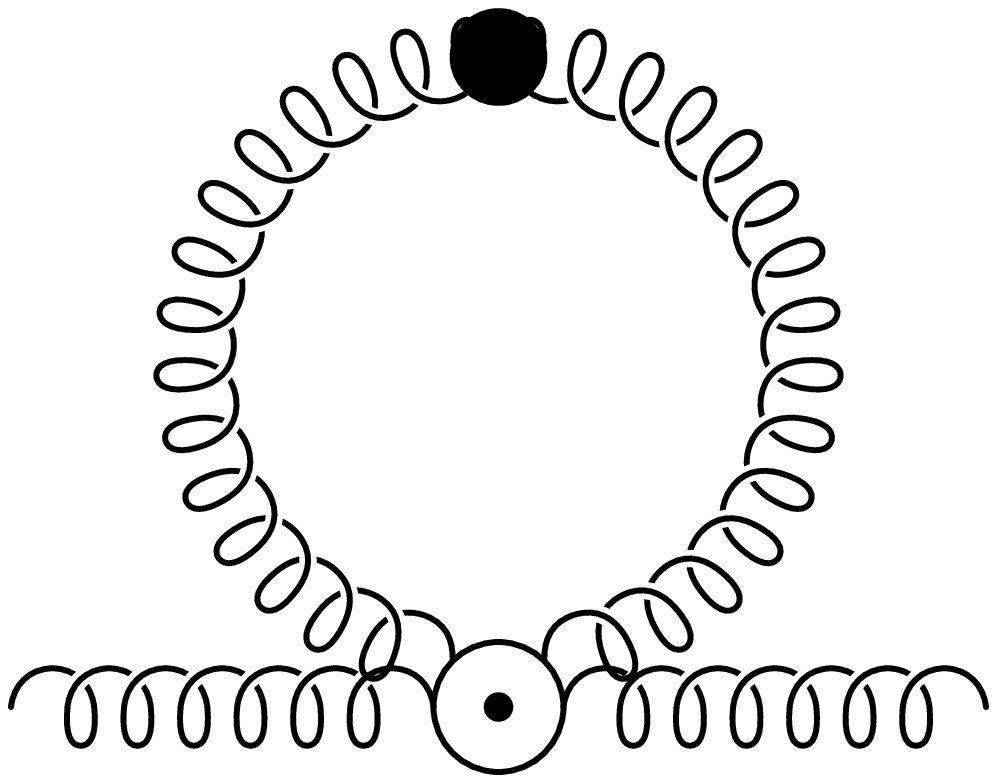}
 }
\\
\parbox{11cm}{
  \caption{\label{fig1}The diagrams contributing to the next-to-leading order
  of the thermal gluon polarization function. The inner momentum is soft, so
  the propagators and vertices are the resummed ones.}
}
\end{center}
\end{figure}

The contributions to $\delta\Pi_{\mu\nu}$ are contained in the diagrams of 
figure \ref{fig1},
\begin{equation}\label{eff220}
\delta \Pi_{\mu\nu} (Q) 
 =  \Pi^{\rm loop}_{\mu\nu}(Q) + \Pi^{\rm tad}_{\mu\nu}(Q)
   -\Pi^{\rm [-]}_{\mu\nu}(Q)
\end{equation}
with
\begin{eqnarray}
\Pi^{\rm loop}_{\mu\nu}(Q)  & = & 
  \frac{1}{2}g^2N{\textstyle\sum}\!\!\!\!\!\!\int\nolimits_P 
   {}^*G^{\rho\sigma}(K) {}^*G^{\lambda\tau}(P)\nonumber\\
&& {}^*\Gamma_{\mu\rho\lambda}(Q,-K,-P)\,{}^*\Gamma_{\nu\sigma\tau}(Q,-K,-P)\;,
\nonumber\\
\Pi^{\rm tad}_{\mu\nu}(Q)  & = & 
  \frac{1}{2}g^2N{\textstyle\sum}\!\!\!\!\!\!\int\nolimits_P 
  {}^*G^{\rho\sigma}(P)\,{}^*\Gamma_{{\mu\nu}\rho\sigma}(Q,-Q,-P,P)
\nonumber
\;.
\end{eqnarray}
The two parts $\Pi^{\rm loop}$ and $\Pi^{\rm tad}$ correspond to the two
diagrams in figure \ref{fig1}.
By subtracting $\Pi^{[-]}$, which is given by figure \ref{fig1} with hard
inner momentum, $\delta\Pi$ is made up of soft inner momenta automatically.
The limit of hard momentum, however, does not necessarily imply to replace the
resummed propagators and vertices with the tree-level ones. Instead, whenever 
the outer momentum $Q$ becomes lightlike, one is forced to use massive 
propagators \cite{eFFacT,LebSmil}. We will cover this case in the following 
section.

\section{Resummation of asymptotic masses}
\label{minf}

For lightlike outer momentum the use of tree-level propagators to obtain 
$\Pi^{[-]}$ in \gl{eff220} leads to an incomplete subtraction of the hard 
momentum scale in the next-to-leading order selfenergy $\delta\Pi$ 
\cite{sed,eFFacT}. Then, strong collinear singularities in $\delta\Pi$ 
spoil the conventional perturbative expansion. 

As an example for the underlying mechanism consider the toy-term
\begin{equation}\label{eff260}
\delta\Pi^{\rm Example}=-4e^2\!\int\!\!{d^3p\over (2\pi)^3}{n_b(p)\over p}\,
 \left\{ {\vec p\vec q\over \omega_p\omega-\vec p\vec q}
       - {\vec p\vec q\over p\omega-\vec p\vec q} \right\}
\end{equation}
with $\omega_p^2=p^2+m^2$.
The first part of $\delta\Pi^{\rm Example}$ is typical for the one-loop
diagrams in figure \ref{fig1}. For simplicity we have replaced the complicated
self-energies by a single mass $m$. The second part is the analogon to
$\Pi^{[-]}$ in the tree-level approximation. 

When $Q$ is off the light-cone, i.e.~$\omega>q$, the leading order of the 
integral in \gl{eff260} obviously comes from soft $p$. We can approximately 
set $n_b(p)\approx T/p$ and evaluate \gl{eff260}. But with $Q$ becoming 
light-like, $\omega\simge q$, the two parts in \gl{eff260} behave totally
different. The second term becomes large due to a collinear singularity
while the first one remains finite. Now, the integration is not any more 
restricted to the soft scale. Still using $n_b(p)\approx T/p$, we find a 
strong power-like singularity \cite{sed},
\begin{eqnarray}
\delta\Pi^{\rm Example} 
 &\to& -{e^2\mu T\over  2\pi}{\omega\over q}\int_{-1}^1\!du\,
      {u^2\over {\sqrt{\omega^2/q^2-u^2}}^3}
 \;\sim\; {1\over\varepsilon}
\label{eff270}\\
&&{\rm with}\quad
\varepsilon^2 = {\omega^2-q^2\over q^2}\;.\nonumber
\end{eqnarray}
Thus, this behavior is not solely due to a collinear singularity, but also to
a latent UV-singularity, which is caused by a premature restriction to
soft momenta. Obviously, if $\varepsilon$ becomes sufficiently small,
$\delta\Pi$ may overgrow the leading order self-energy $\Pi$ and the 
perturbative expansions breaks down.

For the consistency of the Braaten-Pisarski effective action \cite{eff} the 
above behavior is a disaster. The hard scale has not been integrated out 
entirely and consequently the action of \cite{BP} is incomplete in the sense
of renormalization group theory. To restore consistency, as seen in
\cite{eFFacT}, one has to include a certain asymptotic mass
\begin{equation}\label{eff280}
m_\infty^2=\Pi_t(Q^2=0)={g^2NT^2\over 6}
\end{equation}
in the calculation of the hard thermal loops. 
Fortunately, though including $m_\infty$, in the corresponding
improved HTL-effective action \cite{eFFacT} the elegant structure
and gauge-invariance of the Braaten-Pisarski action \cite{eff} can be 
maintained. Moreover, recalculating the gluon polarization function with the
improved action gives finite results for all momenta, including the 
light-cone.

\section{Consistency conditions}
\label{consist}

The true gluon system has a richer structure than in the previous example. 
Here, the above mentioned latent UV-singularities are not the only mechanism
that cause $\delta\Pi$ to grow near the light-cone. As will be seen in the
following section, the HTL-parts of the vertices introduce additional collinear
singularities which may cause a breakdown of the perturbation expansion.
However, the physical quantity under consideration is not the self-energy, but
plasmon frequency and damping. It is shown in this section, that the physical
quantities can tolerate a certain amount of light-cone enhancements in the
self-energy without spoiling their perturbative expansion.

Consider equation \gl{eff210} that defines $\delta\omega_\ell$ and 
$\gamma_\ell$. For its derivation we made use of two (really trivial)
inequalities,
\begin{equation}\label{qcd110}
\delta\omega_\ell,\gamma_\ell\ll\omega_\ell\qquad{\rm and}\qquad
\delta\Pi_{oo}\ll \Pi_{oo}
\end{equation}
for all physical momenta. Violation of the inequalities \gl{qcd110} 
would invalidate the perturbative expansion of the physical quantities. 
In the following section we check \gl{qcd110} by calculating $\delta\Pi_{oo}$ 
near the light-cone. 

However, we will {\em not} do the calculation {\em on} the
light-cone but we will keep a small distance $g\ll\varepsilon\ll 1$ with 
$\varepsilon^2=(\omega^2-q^2)/q^2$ for two reasons:
\begin{itemize}
\item On the light-cone the use of the improved HTL-vertices \cite{eFFacT} is
required, leading into technical difficulties. 
For $\varepsilon\gg g$, however, we may neglect $m_\infty$, which renders the
following calculation feasible.
\item A finite but small distance  $\varepsilon\gg g$ is still interesting to
decide whether the asymptotic mass resummation is sufficient near the 
light-cone. If there were a breakdown of perturbation theory (there is none!) 
we would have to think about an additional resummation to $m_\infty$, for
example one including some kind of thermal damping, see eg~\cite{dampf}. 
Such a new screening mechanism for the collinear singularities may show up in 
a larger (though still small) distance $\varepsilon$, namely 
$\varepsilon^2\sim g$ (rather than $\varepsilon^2\sim g^2$ in the case of 
$m_\infty$). But this range is consistently treated within the present
framework. 
\end{itemize}

Let us return to the conditions \gl{qcd110}. Since in \gl{eff210} we have
$\partial_\omega \Pi_{oo}\sim g^2T^2/\varepsilon^2$ near the light-cone, the 
first condition in \gl{qcd110} is a rather weak one for the self-energy 
$\delta\Pi$. It may behave like
\begin{equation}\label{qcd120}
\Re e\;\delta\Pi_{oo},\;\Im m\,\delta\Pi_{oo} \sim g^3T^2{1\over \varepsilon^2}
\end{equation}
without violating $\delta\omega_\ell,\gamma_\ell\ll\omega_\ell$. Even 
additional logarithms $\ln(\varepsilon)$ are allowed. The second inequality in 
\gl{qcd110} is more restrictive. The leading order $\Pi_{oo}$ behaves like 
$\Pi_{oo}\sim g^2T^2\ln(\varepsilon)$ and thus
\begin{equation}\label{qcd130}
\Re e\,\delta\Pi_{oo} \simle g^3T^2{1\over \varepsilon}
\end{equation}
is allowed as long as $\varepsilon\simge g$.
The  condition \gl{qcd130} does not apply to the imaginary part because the
leading order $\Im m\,\Pi$ is zero for physical momenta.

Combining \gl{qcd120} and \gl{qcd130}, we arrive at the following inequalities
which form the consistency conditions to be fulfilled by $\delta\Pi_{oo}$ for
$g\ll\varepsilon\ll 1$:
\begin{equation}\label{qcd140}
\Re e\,\delta\Pi_{oo} \simle g^3T^2{1\over \varepsilon} 
\quad{\rm and}\quad
\Im m\,\delta\Pi_{oo} \simle g^3T^2{1\over \varepsilon^2} \;.
\end{equation}
These restrictions on the light-cone behavior of $\delta\Pi_{oo}$
will be checked by explicit calculation in the following section.

\section{Plasmon frequency near the light-cone}
\label{eval}

In this main section we evaluate the next-to-leading gluon polarization 
function $\delta\Pi_{oo}(Q)$ near the light-cone $Q^2=0$. We are interested in 
the order of magnitude of the first term of the asymptotics in 
$\varepsilon^2=Q^2/q^2$.

The starting point of the calculation is equation (4.5) of \cite{FleSch}. 
The expression for $\delta\Pi_{oo}$ can be split into two parts,
\begin{equation}\label{qcd100}
  \delta\Pi_{oo}(Q) =: \delta\Pi^{\rm tree}_{oo}(Q)
                     + \delta\Pi^{\rm HTLV}_{oo}(Q) \;. 
\end{equation}
The first part contains only tree level vertices while all the remaining 
contributions from HTL-vertices are assembled in $\delta\Pi^{\rm HTLV}_{oo}$.
In both parts the propagators are the resummed one \gl{eff100}.

The tree-level part $\delta\Pi^{\rm tree}_{oo}$ has been studied in 
\cite{FleSch}. There, one obtains the mechanism discussed in section \ref{minf}
leading to the asymptotic mass resummation. With use of the improved HTL 
effective action \cite{eFFacT} this part will be completely finite near the 
light-cone without including any enhancement effects.

In $\delta\Pi^{\rm HTLV}_{oo}$, however, the HTL-vertices indeed introduce
strong collinear effects\footnote{In fact, this part has been studied in 
\cite{FleSch} as well. But as the present mechanism leading to  
$1/\varepsilon^2$ terms was missed, the result given in \cite{FleSch} remained
incomplete.}.
This part is given by
\begin{equation}\label{eff500}
\delta\Pi_{oo}^{\rm\scriptsize HTLV}(Q)
= \Big\{ \Sigma + \Upsilon + \Phi + \Psi \Big\} 
\end{equation}
with
\begin{eqnarray}
 \Sigma & = & \frac{1}{2} g^2N{\textstyle\sum}\!\!\!\!\!\!\int\nolimits_P
              \Delta \delta\Gamma_{oooo} \;,
\label{eff510}\\
 \Upsilon & = & \frac{1}{2} g^2N {\textstyle\sum}\!\!\!\!\!\!\int\nolimits_P
                \Delta\Delta^-(\delta\Gamma_{ooo})^2 \;,
\label{eff520}\\
 \Phi & = & g^2N{\textstyle\sum}\!\!\!\!\!\!\int\nolimits_P 
            \Delta_t\Delta^-
            \Big\{ \delta\Gamma_{oo\mu}\delta\Gamma_{oo}{}^\mu \nonumber\\
&&\quad\qquad     + 2(P_o+Q_o)\delta\Gamma_{ooo}\Big\} \;,
\label{eff530}\\
 \Psi & = & \frac{1}{2}g^2N{\textstyle\sum}\!\!\!\!\!\!\int\nolimits_P
            \Delta_t\Delta_t^-
            \delta\Gamma_{o{\mu\nu}}\delta\Gamma_o{}^{\mu\nu}
\label{eff540} \;,
\end{eqnarray}
as well as $\Delta = (\Delta_t-\Delta_\ell)P^2/p^2$. The HTL-parts of
the vertices $\delta\Gamma$ are given by \gl{eff160} and \gl{eff170}. Remember,
$Q$ refers to the outer momentum, $P$ is summed over and $K=Q-P$. 
We will concentrate on the terms $\Sigma$ and $\Upsilon$ which are the only
ones giving rise to strong collinear singularities \cite{diss}.

For summing over the Matsubara frequencies it is convenient to introduce
spectral densities $\Delta(P_o,p)=\int\!dx\,\rho(x,p)/(P_o-x)$. The explicit 
form of $\rho$ is given e.g.~in appendix B of \cite{nt}. Since in
$\delta\Pi^{\rm HTLV}_{oo}$ the variables $p$ and $x$ are soft from the outset
we may use $n_b(x)\approx{T\over x}$. 
\begin{eqnarray}
\Sigma & = & g^2NT\int\!\!{d^3p\over (2\pi)^3}\,
      \int\!\!dx\,{\varrho(x,p)\over x}\nonumber\\
&&    \int_\Omega {3m^2\over (YQ)^2}
      {Q_o\over Q_o-x-\vec e\vec k} \;,
\label{eff550}\\
\Upsilon & = & -g^2NT\int\!\!{d^3p\over (2\pi)^3}\,
  \int\!\!dxdy {\varrho(x,p)\varrho(y,k)\over Q_o-x-y} \nonumber\\
&&  \int_{\Omega\Omega'}{3m^2\over YQ}{3m^2\over Y'Q} 
  \left\{ 
    {1\over x}{Q_o-x\over Q_o-x-\vec e\vec k}{Q_o\over Q_o-x-\vec e'\vec k} 
  \right. \nonumber\\
&&\left.
    - 2{Q_0\over Q_o-\vec e\vec p-\vec e'\vec k}{1\over Q_o-x-\vec e'\vec k} 
\right\}\;.
\label{eff560}
\end{eqnarray}
At this stage we are allowed to perform the analytical continuation
$Q_o\to\omega +i\eta$, $\eta\to +0$. In the following we keep with the
notation $Q_o$ but always mean $\omega +i\eta$.

As we will see in the next subsection the main reason for the strong 
light-cone singularities is a combination of collinear and infrared effects. 
This is due to the weird behavior of the transversal density \cite{nt} for
small momentum,
\begin{equation}\label{eff570}
{\varrho_t(x,p)\over x} \to {1\over p^2}\delta(x)
\qquad{\rm for}\qquad p\to 0 \;,
\end{equation}
which is nothing but the lack of a magnetic screening mass. We split off
the IR-sensitive part,
\begin{eqnarray}
\Sigma = \Sigma_o + (\Sigma-\Sigma_o) = \Sigma_o+\Sigma_1\:,
\nonumber\\
\Upsilon = 2\Upsilon_o + (\Upsilon-2\Upsilon_o) = 2\Upsilon_o + \Upsilon_1
\label{eff575}
\end{eqnarray}
where the factor 2 in the second equation is for conveniance. It comes from 
the fact that $\Upsilon$ is invariant with respect to the interchange 
$\vec p\to \vec k$. So, we have two infrared regions, $p\to 0$ as well as 
$k\to 0$. Using \gl{eff570} to neglect $x$, the IR-sensitive parts are given by
\begin{eqnarray}
\Sigma_o 
& = & g^2NT\int\!\!{d^3p\over (2\pi)^3}\,
      \int\!\!dx\,{\varrho(x,p)\over x}\nonumber\\
&&\int_\Omega{3m^2\over (YQ)^2} {Q_o\over Q_o-\vec e\vec k} \;,
\label{eff580}\\
\Upsilon_o & = & -g^2NT\int\!\!{d^3p\over (2\pi)^3}\,
      \int\!\!dxdy\, {\varrho(x,p)\varrho(y,k)\over x\;(Q_o-y)}\nonumber\\
&&      \int_{\Omega\Omega'}{3m^2\over YQ}{3m^2\over Y'Q} 
      {Q_o\over Q_o-\vec e\vec k}{Q_o\over Q_o-\vec e'\vec k}
\;.
\label{eff590}
\end{eqnarray}
In fact, these parts are the only ones showing a strong light-cone
behavior \cite{diss}. We will concentrate on them. The remaining terms
$\Sigma-\Sigma_o$ and $\Upsilon-2\Upsilon_o$ are of no relevance in the 
present context.

\subsection{Evaluation of $\Sigma_o$}

We start with the simpler one of the two contributing parts. In \gl{eff580}
the $x$-integration is performed using sum-rules derived eg.~in \cite{rho,nt}.
\begin{equation}\label{meg130}
 \Sigma_o = g^2NT\int\!\!{d^3p\over (2\pi)^3} 
            {1\over p^2}{3m^2\over  p^2+3m^2} 
            \int_\Omega{3m^2\over (YQ)^2} {Q_o\over YQ+\vec e\vec p}
\end{equation}
with rewriting $Q_o-\vec e\vec k = YQ+\vec e\vec p$. Near the light-cone 
$Q^2=0$ the denominator $(YQ)^2$ tends to diverge. At $YQ=Q_o-\vec e \vec q=0$
it gives rise for a powerlike collinear singularity. In addition, the second
denominator in \gl{meg130}, $YQ+\vec e\vec p$, mixes the collinear point $YQ=0$
with the infrared region of the $p$ integration. For vanishing $YQ$ we obtain
a  logarithmic singularity at small $p$, for vanishing $p$ the collinear 
singularity becomes stronger. So, from \gl{meg130} we expect the leading 
contribution to come from a very limited region of the integration variables: 
the infrared limit of the $p$-Integral.

More precisely, after performing the angular integration, we have
\begin{eqnarray}
\lefteqn{ \Sigma_o =  g^2NT\int\!\!{d^3p\over (2\pi)^3} 
                 {1\over p^2}{3m^2\over  p^2+3m^2} 
                 {m_\infty^2\omega^2(\vec p\vec q - p^2) 
                  \over  {\sqrt{(\vec p\vec q)^2 + Q^2p^2}}^3} 
        }\nonumber\\
&& 
   \left\{\ln\left({Q^2+\vec p\vec q + \sqrt{\phantom{x}} 
              \over  Q^2+\vec p\vec q - \sqrt{\phantom{x}} }\right) 
   - i\pi\Theta\left(p^2-2\vec p\vec q-Q^2\right)\right\}\;. 
\label{meg133}
\end{eqnarray}
Indeed, $Q^2$ regulates the infrared behavior of the $p$-integral, as we can 
see in particular in the step-function of the imaginary part of \gl{meg133}.
The leading contribution of the light-cone asymptotics comes from $p^2$ and 
$\vec p\vec q$ becoming comparable small to $Q^2$.

Accordingly we scale $p\to \varepsilon qv$ and $\vec p\vec q/pq \to 
\varepsilon u$. Now the integration variables are dimensionless.
\begin{eqnarray}
\Sigma_o &=&  {1\over \varepsilon^2}\Gamma\int_0^\infty\!\!dv
  \int_{-1/\varepsilon}^{1/\varepsilon}\!\!du 
  {\alpha^2\over v^2+\alpha^2} {u-v\over v^2{\sqrt{1+u^2}}^3}
\nonumber\\
&&
  \left\{\ln\left({{1\over v}+u+\sqrt{\phantom{x}}
              \over {1\over v}+u-\sqrt{\phantom{x}} }\right)
  -i\pi\Theta\left(v^2-2uv-1\right) \right\}
\label{meg135}
\end{eqnarray}
with $\Gamma={g^2NTm_\infty^2/(4\pi^2q)}$ and 
$\alpha^2={3m^2/(\varepsilon^2q^2)}$. 
The latter constant $\alpha$, becomes large near the light-cone. However, 
we are not allowed to perform $\alpha\to\infty$ since a finite $\alpha$ is 
needed to control the $v$-integral for large values of $v$. Nevertheless, 
evaluating \gl{meg135} analytically,
\begin{eqnarray}
\Im m\,\Sigma_o 
& = & \pi\Gamma\left( \ln(\alpha^2/\rho) -2 +{\cal O}(\varepsilon^2)  \right)
\nonumber\\
& = &  \pi\Gamma\left( \ln({12m^2 \over  q^2\varepsilon^4}) 
                       - 2 +{\cal O}(\varepsilon^2)\right) \;,
\label{meg140}\\
\Re e\,\Sigma_o 
 & = & {1\over \varepsilon^2}\Gamma \left( -\pi^2 
                                   + {\cal O}(\varepsilon^2) \right) \;,
\label{meg150}
\end{eqnarray}
with $\rho=(\sqrt{1+\varepsilon^2}-1)/(\sqrt{1+\varepsilon^2}+1)
\sim \varepsilon^2/ 4$. Indeed, we obtain strong collinear singularities 
$\sim\varepsilon^{-2}$. For the real part, the result \gl{meg150} violates, 
if considered separately, the inequality \gl{qcd140} given in the last 
section. Thus, assuming \gl{meg150} is the final result, the perturbative 
expansion of the physical quantities $\omega_\ell$ and $\gamma_\ell$ would 
be broken.

\subsection{Evaluation of $\Upsilon_o$}

Compared to $\Sigma_o$ the evaluation of $\Upsilon_o$ is more involved 
due to the additional propagator $\Delta^-$ in \gl{eff520}. Before we discuss
simplifications for $\Delta^-$ we turn to \gl{eff590}, apply the sum rules and
perform the angular integrations.
\begin{eqnarray}
\lefteqn{
\Upsilon_o = \frac{1}{2} g^2NT\int\!\!{d^3p\over (2\pi)^3}
   {1\over p^2}{3m^2\over p^2+3m^2} 
   {\Delta(Q_o,k) m_\infty^4\omega^2
    \over \sqrt{(\vec p\vec q)^2+\varepsilon^2p^2q^2}{}^2}
}
\nonumber\\
&& \left\{\ln\left({Q^2+\vec p\vec q + \sqrt{\phantom{x}}
               \over Q^2+\vec p\vec q - \sqrt{\phantom{x}} }\right)
   -i\pi\Theta\left(p^2-2\vec p\vec q - Q^2\right)  \right\}^{\displaystyle 2}
\label{meg160}\;.
\end{eqnarray}
Despite of the additional propagator, \gl{meg160} has the same structure than 
$\Sigma_o$ in \gl{meg130}. Indeed, the leading $\varepsilon$-order of 
$\Upsilon_o$ comes from the same very limited region of the integration 
variables as in $\Sigma_o$: $p^2$ as well as $\vec p\vec q$ are comparable 
small to $Q^2$. This opens the possibility for suitable approximations in 
the propagator $\Delta^-$.

\subsubsection{The propagator}

The propagator $\Delta$ in \gl{meg160} is made up of two parts,
\begin{eqnarray}
\Delta(Q_o,k) 
 &=& {1\over k^2}\left\{ {Q_o^2-k^2\over Q_o^2-k^2-\Pi_t} \right. \nonumber\\
&& \left.              - {Q_o^2-k^2\over (Q_o^2-k^2)(1+\Pi_{oo}/k^2)}  \right\}
\;. \label{meg180} 
\end{eqnarray}
The transversal part (first term) does not contribute to the leading order 
since its numerator behaves like $Q^2$ and in the denominator $\Pi_t$ does 
not vanish for $Q_o\approx k$. Hence,
\begin{equation}\label{meg190}
\Delta(Q_o,k)\approx
 { -1 \over  k^2 + 3m^2\left[1-\frac{1}{2}{Q_o\over k}
                               \ln\left({Q_o+k\over Q_o-k}\right) \right] } \;.
\end{equation}
When we do the analytical continuation $Q_o\to \omega +i\eta$ ($\eta\to +0$) 
the denominator has besides the common \cite{FRoSch} $i\sigma\eta$, 
$\sigma=\pm 1$ a further imaginary part from the cut of the logarithm,
\begin{equation}\label{meg200}
\ln\left({Q_o+k\over Q_o-k}\right) 
\to\ln\left|{\omega+k\over \omega-k}\right|-i\pi\Theta(k^2-\omega^2)
\;.
\end{equation}
Thus
$$
\Delta(\omega,k) \approx {-1 \over  k^2+ \Pi_{oo}(\omega,k) 
 + i\pi m_\infty^2{\omega\over k}\Theta(k^2-\omega^2) + i\eta\sigma } \;. 
$$
With regard to the value of the step function $\Theta$ we distinguish two 
cases.

\noindent
\underline{\boldmath$\Theta=0$\unboldmath} ($k^2<\omega^2$). In this case
\begin{eqnarray}
\Delta_0 & = & {-1 \over  k^2+\Pi_{oo}(\omega,k)} 
                        + i\pi\sigma \delta\left(k^2+\Pi_{oo}\right) \\
 & = & {-1 \over  k^2+\Pi_{oo}(\omega,k)} 
  + i\pi\sigma{ \delta(k^2-q^2) 
               \over  |1 + \partial_{q^2}\Pi_{oo}(\omega,q)|} \;.
\label{meg220}
\end{eqnarray}
Near the light-cone we have $\sigma=1$ and $\partial_{q^2}\Pi_{oo}
=-m_\infty^2/Q^2$. Remember, we have to evaluate $\delta\Pi$ on the 
longitudinal mass-shell, see \gl{eff210}. Thus $\omega=\omega_\ell(q)$
is given and $k^2 + \Pi_{oo}(\omega,k)$ is zero at $k^2=q^2$, because this 
is the longitudinal mass-shell condition defining $\omega_\ell$.

In the term of \gl{meg220} with the delta-function we will find the known
problems with the mass shell singularity. Here the only true IR-divergence
will occur following the mechanism described in \cite{FRoSch}. We will
regulate this by including an artificial small cut-off mass $\mu$. All the 
other parts of $\Upsilon$ are IR-finite and do not need any $\mu$.

\noindent
\underline{\boldmath $\Theta=1$\unboldmath} ($k^2>\omega^2$). In this case we 
are allowed to neglect $i\eta$ in the denominator,
\begin{equation}\label{meg230}
\Delta_1 
 = {-1 \over  k^2+\Pi_{oo}(\omega,k) +i\pi m_\infty^2 {\omega\over k} } \;.
\end{equation}

In both cases we simplify $k^2+\Pi_{oo}$ using $q^2,m^2 \gg p^2-2\vec p\vec q$.
As mentioned above we must avoid the use of $Q^2\gg p^2-2\vec p\vec q$. 
In addition we take advantage of being on the longitudinal mass-shell
$ q^2 + 3m^2\left[1-\frac{1}{2}{\omega\over q}
        \ln\left({\omega+q\over \omega-q}\right)\right] = 0$. 
We obtain
\begin{eqnarray}
\lefteqn{k^2 + \Pi_{oo}(\omega,k) }\nonumber\\
 & = & (\vec q-\vec p)^2 
+ 3m^2\left[1-{\omega/2\over \sqrt{(\vec q-\vec p)^2}}
              \ln\left({\omega+\sqrt{\phantom{x}}
                  \over \omega-\sqrt{\phantom{x}}}\right)
      \right]
\nonumber\\
 & \approx&  {3\over 2} m^2\ln\left({\omega^2-k^2\over Q^2}\right) \;.
\label{meg240}
\end{eqnarray}
This simplification, to be used in \gl{meg220} and \gl{meg230}, makes the
following integrals feasible.

\subsubsection{\boldmath $\Theta=0$ part of $\Upsilon_o$}
\label{FROSCH}

We turn to the evaluation of $\Upsilon_o$. For convenience the distinction of 
the two cases $\Theta=0$ and $\Theta=1$ will be retained. Accordingly we split
$\Upsilon_o$ into two parts
\begin{equation}
\Upsilon_o = \Upsilon_o^{\Theta=0} + \Upsilon_o^{\Theta=1} \;.
\end{equation}
Both parts are now evaluated separately.

The part with vanishing step-function is given by
\begin{eqnarray}
\Upsilon_o^{\Theta=0} & = & \frac{1}{2} g^2NT\int\!\!{d^3p\over (2\pi)^3}
{1\over p^2}{3m^2\over p^2+3m^2}
{m_\infty^4\omega^2
 \over \sqrt{(\vec p\vec q)^2+\varepsilon^2p^2q^2}^{\displaystyle 2}}
\nonumber\\
&&\ln^2\left({Q^2+\vec p\vec q + \sqrt{\phantom{x}}
      \over Q^2+\vec p\vec q - \sqrt{\phantom{x}}}\right) 
  \Theta(\omega^2-k^2)
\nonumber\\
&&  \left\{{-1\over m_\infty^2\ln\left({\omega^2-k^2\over Q^2}\right)}
    +{i\pi\sigma\delta(k^2-q^2)\over [1+\partial_q^2\Pi_{oo}(\omega,q)]}  
    \right\}
\;.
\label{meg250}
\end{eqnarray}
First consider the real part of \gl{meg250}.
Scaling $p\to\varepsilon qv$ and $\vec p\vec q/pq \to \varepsilon u$ we have 
\begin{eqnarray}
\lefteqn{
 \Re e\; \Upsilon_o^{\Theta=0}
  = {\Gamma\over 2\varepsilon^2}\int_o^\infty\!\!dv {1\over v^2} 
   {\alpha^2\over  v^2+\alpha^2}
   \int_{-1/\varepsilon}^{1/\varepsilon}\!\!du {1\over u^2+1}
 } \nonumber\\
&& {\Theta(1+2uv-v^2)\over 
    \ln(1+2uv-v^2)}\ln^2\left({1/v+u+\sqrt{u^2+1}
                        \over  1/v+u-\sqrt{u^2+1}} \right)
\label{meg255}
\end{eqnarray}
where $\Gamma={g^2NTm_\infty^2/(4\pi^2 q)}$ and 
$\alpha^2={3m^2/(\varepsilon^2q^2)}$ are the notations we used in \gl{meg135}.
Contrary to \gl{meg135}, the $v$-integral now is finite at both limits small 
and large $v$. For large $v$ it is cut by the step-function. So, here we are 
allowed to perform $\alpha^2\to\infty$. The fact that \gl{meg255} is well
behaved at $v=0$ is seen by expanding the logarithms in $v^2$ and $uv$.

By substituting $v \to \sqrt{1+u^2}v-u$ and $t=u/\sqrt{\phantom{x}}$, 
symmetrizing in $t$ and $v$ as well as introducing $x=(1-v)/(1+v)$ and 
$y=(1-t)/(1+t)$, we obtain
\begin{eqnarray}
 \Re e\,\Upsilon_o^{\Theta=0} 
 &=&  -{1\over 2\varepsilon^2}\Gamma\int_0^1\!\!dx\int_\rho^1\!\!dy 
     {1\over (x-y)^2(1-xy)^2}
\nonumber\\
&&   {(x-y)^2\ln^2(xy)+(1-xy)^2\ln^2(x/y)
    \over  \ln(x/y)-2\ln\left({1+x\over 1+y}\right)} 
\end{eqnarray}
with $\rho=(\sqrt{1+\varepsilon^2}-1)/( \sqrt{1+\varepsilon^2}+1)
\approx\varepsilon^2/4$. However, the integrand $F(x,y)$ is singular at the 
origin $(x,y)=0$ and we are not allowed to neglect $\rho$ here. Instead, we 
exploit $F(x,y)$ being antisymmetric with respect to interchange of $x$ and 
$y$, $F(x,y)=-F(y,x)$. Hence,
\begin{equation}
\Re e\,\Upsilon_o^{\Theta=0} 
= {1\over 2\varepsilon^2}\Gamma\int_\rho^1\!\!dx\int_o^\rho\!\!dy \,F(x,y)\;.
\end{equation}
After scaling both integration-variables $x$ and $y$ with $\rho$ it is finally
possible to neglect $\rho\sim\varepsilon^2$ in the remaining integral. After another change of variables $x\to 1/x$ we obtain the result for the real part of 
$\Upsilon_o^{\Theta=0}$,
\begin{eqnarray}
\Re e\,\Upsilon_o^{\Theta=0}
& =& -{1\over 2\varepsilon^2}\Gamma\int_o^1\!\!dx\int_o^1\!\!dy
    {\ln(xy)\over (1-xy)^2} \nonumber\\
& =& {1\over \varepsilon^2}\Gamma\;{\pi^2\over 6}\;.
\label{meg260}
\end{eqnarray}
Obviously, we have found one more term $\sim\varepsilon^{-2}$ in the real part
of $\delta\Pi_{oo}$ which, if considered separately, would spoil the common
perturbation theory.

We turn to the imaginary part of $\Upsilon_o^0$. Here we find the only true
infrared problem due to a mass-shell singularity \cite{FRoSch}.
It needs an artificial infrared regulator in the transversal part
of the propagator, $\Delta_t(P_o=0)=-(\vec p^2+\mu^2)^{-1}$. Consider the mass 
$\mu$ to be small compared to the soft scale ($\mu\ll gT$) as well as 
$\mu/\varepsilon\ll gT$. With the latter condition we again restrict ourself 
to a distance $\varepsilon$ with $g\ll\varepsilon\ll 1$. While in section 5 we
used this restrictions to neglect the asymptotic mass in the HTL- propagators
and -vertices (otherwise the present calculation would have been undoable) we 
now need it to solve the integral in \gl{meg265}.

In $\Im m\,\Upsilon_o^{\Theta=0}$ we first exploit the delta-function in 
\gl{meg250}. Then we proceed as with the real part. Scaling 
$p\to\varepsilon qv$ and with $\alpha^2\to\infty$ we obtain
\begin{eqnarray}
\Im m\, \Upsilon_o^{\Theta=0} 
& =&  {g^2NT\over (2\pi)^2}{m_\infty^2\omega^2\over \varepsilon^2q^3}
   {\pi\sigma m_\infty^2\over \varepsilon^2q^2|1+\partial_q^2\Pi_{oo}|}
   {1\over 4}\int_0^{1/\varepsilon}\!\!dv \nonumber\\
&&     {1\over v}{1\over v^2+1}{1\over v^2+b^2}
     \ln^2\left({\sqrt{v^2+1}+v\over \sqrt{v^2+1}-v}\right)
\label{meg265}
\end{eqnarray}
where the dependence on the artificial infrared cutoff $\mu$ is contained in
$b^2=\mu^2/(4q^2\varepsilon^2)$. While $b$ is small, see above, the 
$b$-asymptotics of the $v$-integral can be calculated analytically.
We finally obtain for the imaginary part of $\Upsilon_o^{\Theta=0}$
\begin{equation}\label{meg270}
 \Im m\,\Upsilon_o^{\Theta=0} 
= -\pi{\Gamma\over 2\varepsilon^2}\left\{ 
     \ln\left(\mu^2\over q^2\varepsilon^2\right) +{7\over 4}\zeta(3)-3
   \right\}
\;.
\end{equation}
The logarithm, containing the $\mu$-dependence, is known from several
work on plasmon-damping \cite{FRoSch}.

\subsubsection{\boldmath $\Theta=1$ part of $\Upsilon_o$}

With the propagator $\Delta_1$ from \gl{meg230} our starting point reads
\begin{eqnarray}
\lefteqn{
  \Upsilon_o^{\Theta=1}  = \frac{1}{2} g^2NT\int\!\!{d^3p\over (2\pi)^3}
  {1\over p^2}{3m^2\over p^2+3m^2}
  {m_\infty^2\omega^2
   \over \sqrt{(\vec p\vec q)^2+\varepsilon^2p^2q^2}{}^{\displaystyle 2}}
}\nonumber\\  
&&{-\Theta(k^2-\omega^2) 
   \over  \ln\left({\omega^2-k^2\over Q^2}\right)+i\pi\omega/k}
  \left\{
   \ln\left({Q^2+\vec p\vec q +\sqrt{\phantom{x}}
       \over Q^2+\vec p\vec q-\sqrt{\phantom{x}}}\right)
   -i\pi\right\}^{\displaystyle 2}
\!\!.
\label{meg280}
\end{eqnarray}
Following the transformations and substitutions known from evaluating
$\Re e\,\Upsilon_o^{\Theta=0}$, we obtain
\begin{eqnarray}
\Upsilon_o^{\Theta=1} 
  &=& -{\Gamma\over 2\varepsilon^2}\int_0^1\!\!dx\int_\rho^1\!\!dy 
   {1\over {\cal L}+i\pi}
\nonumber\\
&&   \left\{
   { (\ln(xy)+i\pi)^2\over (1+xy)^2}
    +{\left(\ln\left({x\over y}\right)+i\pi\right)^2\over (x+y)^2}
   \right\}
\label{meg290}
\end{eqnarray}
where ${\cal L} = \ln\left({x(1+y)^2\over y(1-x)^2}\right)$. Splitting in real
 and imaginary part,
\begin{equation}
\Upsilon_o^{\Theta=1} 
 = -{\Gamma\over 2\varepsilon^2}\int_0^1\!\!dx\int_\rho^1\!\!dy\,
    { F_R(x,y) - i\pi F_I(x,y)\over {\cal L}^2+\pi^2}
\;,
\end{equation}
\begin{eqnarray}
F_R(x,y) & = &
  {\left(\ln^2(xy)-\pi^2\right){\cal L}+2\pi^2\ln(xy)\over (1+xy)^2}
\nonumber\\
&& +{\left(\ln^2\left({x\over y}\right)-\pi^2\right){\cal L}
   +2\pi^2\ln\left({x\over y}\right)
   \over (x+y)^2}
\;,\\
F_I(x,y) & = & 
  {\ln^2(xy)-2{\cal L}\ln(xy)-\pi^2\over (1+xy)^2}
\nonumber\\
&&+{\ln^2\left({x\over y}\right)-2{\cal L}\ln\left({x\over y}\right)-\pi^2
    \over (x+y)^2}
\;.
\end{eqnarray}
As both functions $F(x,y)$ are singular at the origin $(x,y)=0$, we are not 
allowed to neglect $\rho\sim\varepsilon^2$. Instead, we isolate the singular
parts (index \$),
\begin{eqnarray}
F_R^\$(x,y) &=& {\ln\left({x\over y}\right)({\cal L}^2+\pi^2) \over  (x+y)^2} 
\;,\label{meg300}\\
F_I^\$(x,y) &=& -{{\cal L}^2+\pi^2 \over  (x+y)^2}
\;.\label{meg305}
\end{eqnarray}
These parts can easily be integrated over,
\begin{eqnarray}
\int_0^1\!\!dx\int_\rho^1\!\!dy\,{\ln\left({x\over y}\right)\over  (x+y)^2}
  &=& -{\pi^2\over 6} + {\cal O}(\rho\ln(\rho))
\;,\\
\int_0^1\!\!dx\int_\rho^1\!\!dy\,{-1\over  (x+y)^2}
 &=& \ln(\rho)+\ln(2) +{\cal O}(\rho)\;.
\end{eqnarray}
In the remaining integrals the differences $F-F^\$$ are regular on the entire
$(x,y)$-interval. So, for the leading term of the $\varepsilon$-asymptotic we
are now allowed to set $\rho=0$. Nevertheless, the integrals are not trivial
at all\footnote{Thanks to Hermann Schulz and Anton Rebhan for their  
support in calculating these integrals}. For technical details we refer to
the appendix. The result of the calculation is,
\begin{eqnarray}
\int_0^1\!\!dx\,dy\,
 { F_R(x,y) - F_R^\$(x,y)\over {\cal L}^2+\pi^2}
& = &  - {\pi^2\over 2}\;,
\label{meg310}\\
\int_0^1\!\!dx\,dy\,
    { F_I(x,y) - F_I^\$(x,y)\over {\cal L}^2+\pi^2}
& = & {7\over 4}\zeta(3)-\ln(2)-1 
\;.
\label{meg320}
\end{eqnarray}

Combining these parts to $\Upsilon_o^{\Theta=1}$ we find for the real and 
imaginary part
\begin{eqnarray}
\Re e\, \Upsilon_o^{\Theta=1} 
 = -{\Gamma\over 2\varepsilon^2}\left\{\;, -{2\pi^2\over 3} \;,\right\} 
\;,
\label{meg330}\\
\Im m\, \Upsilon_o^{\Theta=1}
 = \pi{\Gamma\over 2\varepsilon^2}
   \left\{ \ln(\rho) + {7\over 4}\zeta(3) -1 \right\}\;.
\label{meg340}
\end{eqnarray}
Again, there are unwanted strongly growing terms $\sim \varepsilon^{-2}$ in 
the real part.

\section{Result and conclusion}
We have considered the next-to-leading order of the plasmon self-energy
$\delta\Pi_{oo}$ near the light-cone. Evaluating the leading order term 
of an asymptotic expansion of $\delta\Pi_{oo}$ in $\varepsilon^2=Q^2/q^2$, 
we obtain several contributions that give rise to collinear singularities 
with a strength that would spoil the perturbative expansion of the physical 
quantities, even if the improved HTL-resummation  \cite{eFFacT} is used.
However, the final result is the sum of the parts obtained in \gl{meg140},
\gl{meg150} as well as twice the ones found in \gl{meg260}, \gl{meg270},
\gl{meg330} and \gl{meg340},
\begin{eqnarray}
\Re e\,\delta\Pi_{oo}
 &=&  {g^2mNT\over \pi}\left\{ {1\over \varepsilon^2}\cdot 0 
                              +{\cal O}({1\over\varepsilon})\right\} \;,
\label{meg350}\\
\Im m\,\delta\Pi_{oo}
 &=&  {g^2mNT\over \pi}
   \left\{ {1\over \varepsilon^2}{3m\over 8q}\ln\left({3m^2\over \mu^2}\right)
           +{\cal O}({1\over\varepsilon})\right\}
 \,.\label{meg360}
\end{eqnarray}
In the real part, all the unwanted terms $\sim \varepsilon^2$ precisely
cancel out. Hence, the consistency conditions of section \ref{consist} are
indeed fulfilled. The perturbative expansion of the physical quantities
$\omega_\ell$ and $\gamma_\ell$ using the improved HTL-action \cite{eFFacT} 
is valid, even for momenta near the light-cone. The positive result of the 
present investigation shows no need for further improvements of the  
Braaten-Pisarski scheme for light-like momenta besides the one based on the 
asymptotic mass.

The true leading order term of the real part behaves $\sim g^3T^2/\varepsilon$.
This term is non-vanishing. For its derivation  we refer to \cite{diss} and 
\cite{FleSch}. However, as discussed in section \ref{consist}, such a term
does not break the consistency of the perturbative expansion.

Nevertheless, in calculating other physical quantities with light-like
momenta, further corrections to the effective action may well contribute. For
example for the soft real photon production rate \cite{srppr} one has to 
evaluate the imaginary part of the photon self-energy which turns out to be 
$\sim e^2g^3T^2$.  Hence, compared to our present result $\sim g^3T^2$, 
it is down by two powers of a coupling constant. 
It is well known, for calculations with high precision one has to
use effective actions that include higher order interaction terms \cite{BN}.
So, corrections to the HTL effective action, eg.~some new kind of
plasmon interaction-term, would contribute in evaluating higher order 
quantities. But, as shown in the present analysis, such a term must not be 
resummed into  the leading order of the soft plasmon propagator. Instead it 
would be a  common perturbative correction\footnote{A recent paper by 
Petitgirard \cite{g} supports this conjecture}.

\section*{Acknowledgment}

I am grateful to Anton Rebhan and Hermann Schulz for valuable discussions
and ideas as well as their help in calculating the integrals.
I would like to thank Emmanuel  Petitgirard for discussions during his 
visit in Hannover and Patrick Aurenche and the LAPTH for their hospitality
during a three month stay in Annecy.

%
%
\begin{appendix}

\section{Integrals}

\label{ANHC}
In the maintext the real and imaginary part of $\Upsilon_o^1$ were traced back
to the integrals \gl{meg310} and \gl{meg320}. Here we treat the second, which
is
\begin{eqnarray}
{\cal K} &=& \int_0^1\!\!dx\int_0^1\!\!dy\,{1\over {\cal L}^2+\pi^2}
           \left\{ F_I(x,y) - F_I^\$(x,y)\right\}
\nonumber\\
 &=& {7\over 4} \zeta(3) -\ln(2) -1 \label{pla110}\;.
\end{eqnarray}
The other integral \gl{meg310} can be evaluated in an analogous manner. In 
\gl{pla110}  ${\cal L} = \ln\left({x\over y}{(1+y)^2\over (1-x)^2}\right)$ and
\begin{eqnarray}
\lefteqn{F_I(x,y) - F_I^\$(x,y) }
\nonumber\\
 &=& {(\ln(xy)-{\cal L})^2-({\cal L}^2+\pi^2)\over (1+xy)^2}
  + {\left(\ln\left({x\over y}\right)-{\cal L}\right)^2\over (x+y)^2 } \; .
\label{pla120}
\end{eqnarray}
With \gl{pla120} in \gl{pla110} we may split 
${\cal K} = {\cal K}_o + {\cal K}_1$ with
\begin{eqnarray}
{\cal K}_o & = & -\int_0^1\!\!dx\int_0^1\!\!dy\,{1\over (1+xy)^2}
 = -\ln(2) \;,\\
\label{pla130}
{\cal K}_1 & = & \int_0^1\!\!dx\int_0^1\!\!dy\,{1\over {\cal L}^2+\pi^2}
\nonumber\\
&& \left\{ {(\ln(xy)-{\cal L})^2\over (1+xy)^2}
          +{\left(\ln\left({x\over y}\right)-{\cal L}\right)^2\over (x+y)^2 }
   \right\}
\;.
\label{pla140}
\end{eqnarray}
So, there remains to show that ${\cal K}_1 = {7\over 4}\zeta(3)-1$. We change 
variables in the first term of \gl{pla140} by $y\to 1/y$ and proceed with 
$y\to y/x$. Then, both parts of ${\cal K}_1$ can be joined.
\begin{equation}\label{pla150}
{\cal K}_1 = \int_0^1\!\!dx\,{1\over x}\int_0^\infty\!\!dy\,{1\over (1+y)^2}
{4\ln^2\left({1\over y}{x+y\over 1-x}\right)
 \over \ln^2\left({1\over y}{(x+y)^2\over (1-x)^2}\right)+\pi^2}
\;.
\end{equation}
To simplify the arguments of the logarithms we substitute
$x \to v={1\over y}{x+y\over 1-x}$ and subsequently $y\to {1\over v^2}y$. 
Finally $v\to 1/v$ and $y\to 1/y$ leads to
\begin{equation}
{\cal K}_1 = \int_0^1\!\!dv\,{4\ln^2(v)\over (1-v)^2}\int_0^\infty\!\!dy
{ 1/(y+v^2)+1/(y+v) \over \ln^2(y)+\pi^2}
\,.\!
\end{equation}
To get rid of the logarithms we substitute $\tau=\ln(y)$ and $x=\ln(v)$. Then,
\begin{eqnarray}
{\cal K}_1 &=& \int_o^\infty\!\!dx\,{4e^{-x}x^2 \over  (1-e^{-x})^2}
  \int_{-\infty}^\infty\!\!d\tau\,{1\over \tau^2+\pi^2}
\nonumber\\
 && \quad
   \left({e^\tau\over e^\tau+e^{-2x}} - {e^\tau \over  e^\tau+e^{-x}}\right)
\;.
\end{eqnarray}
The $\tau$-integral may be closed in the upper complex $\tau$-plane. The 
contour runs around the following poles:
\begin{equation}\label{pla170}
 \begin{array}{lcl}
  \tau = i\pi \quad{\rm with\ residue} &:& 
       \displaystyle {1\over 2\pi i}\left({1\over 1-e^{-2x}} 
                                           - {1\over  1-e^{-x}}\right)\;,\\
  \tau = -2x+(2n-1)i\pi &:& 
       \displaystyle{1\over [2x-(2n-1)i\pi]^2+\pi^2}\;, \\
  \tau = -x+(2n-1)i\pi  &:& 
       \displaystyle{-1\over [x-(2n-1)i\pi]^2+\pi^2} \;,
 \end{array}\hspace{-6ex}
\end{equation}
with $n=1,2,\dots,\infty$. In the corresponding sum over $n$ nearly
all terms cancel. The only remaining contributions are
\begin{equation}\label{pla180}
{\cal K}_1 =  \int_o^\infty\!\!dx\,{4e^{-x}x^2 \over  (1-e^{-x})^2}
 \left\{ {1\over 1-e^{-2x}} - {1\over 1-e^{-x}} + {1\over 2x}\right\}\; .
\end{equation}
By splitting in partial fractions
\begin{eqnarray}
{1\over 1-e^{-2x}} & = & 
  \frac{1}{2}\left({1\over 1-e^{-x}}+{1\over 1+e^{-x}}\right) \;, \nonumber\\
{1\over (1-e^{-x})^2}{1\over 1+e^{-x}} & = &
  {1\over 2}\left({1\over (1-e^{-x})^2}+{1\over 1-e^{-2x}}\right)
\nonumber \;,
\end{eqnarray}
and using
\begin{eqnarray}
{1\over 1-e^{-x}}& =& \sum_{m=0}^\infty e^{-mx}\;,\\
{2e^{-x}\over (1-e^{-x})^3} &=& \sum_{m=2}^\infty m(m-1)e^{-mx}\;,
\end{eqnarray}
\gl{pla180} becomes the desired value ${\cal K}_1 = {7\over 4}\zeta(3)-1$.
Hence, \gl{pla110} is derived, as is \gl{meg320} in the maintext.

\end{appendix}

\end{document}